%%%%%%%%%%%%%%%%%%%%%%%%%%%%%%%%%%%%%%%%%%%%%%%%%%%%%%%%%%%%%
%%%These are the macros for submission of papers to hep-th%%%
%%%The default setting is 12pt and 1 page/side but in the%%%%
%%%future it may allow people to choose also 10 pt and%%%%%%%
%%%2 pages/side.%%%%%%%%%%%%%%%%%%%%%%%%%%%%%%%%%%%%%%%%%%%%%
%%%%%%%%%%%%%%%%%%%%%%%%%%%%%%%%%%%%%%%%%%%%%%%%%%%%%%%%%%%%%
%
\def\unlockat{\catcode`\@=11}
\def\lockat{\catcode`\@=12}
\unlockat
\def\d@f@ult{} \newif\ifamsfonts \newif\ifafour
\nonstopmode
%
%%%%%%%%%%%%%%%%%%%%%%
%%%Font definitions%%%
%%%%%%%%%%%%%%%%%%%%%%
%

\font\twelverm=cmr12
\font\ninerm=cmr9
\font\sixrm=cmr6
\font\fourteenbf=cmbx12 scaled\magstep1
\font\twelvebf=cmbx12
\font\ninebf=cmbx9
\font\sixbf=cmbx6
\font\fourteeni=cmmi12 scaled\magstep1      \skewchar\fourteeni='177
\font\twelvei=cmmi12                        \skewchar\twelvei='177
\font\ninei=cmmi9                           \skewchar\ninei='177
\font\sixi=cmmi6                            \skewchar\sixi='177
\font\fourteensy=cmsy10 scaled\magstep2     \skewchar\fourteensy='60
\font\twelvesy=cmsy10 scaled\magstep1       \skewchar\twelvesy='60
\font\ninesy=cmsy9                          \skewchar\ninesy='60
\font\sixsy=cmsy6                           \skewchar\sixsy='60
\font\fourteenex=cmex10 scaled\magstep2
\font\twelveex=cmex10 scaled\magstep1

\ifamsfonts
   \font\ninex=cmex9
   
   \font\sixex=cmex7 at 6pt
   
\else
   \font\ninex=cmex10 at 9pt
   
   \font\sixex=cmex10 at 6pt
   
\fi
\font\fourteensl=cmsl10 scaled\magstep2
\font\twelvesl=cmsl10 scaled\magstep1

\font\sevensl=cmsl10 at 7pt
\font\sixsl=cmsl10 at 6pt

\font\fourteenit=cmti12 scaled\magstep1
\font\twelveit=cmti12

\font\fourteentt=cmtt12 scaled\magstep1
\font\twelvett=cmtt12
\font\fourteencp=cmcsc10 scaled\magstep2
\font\twelvecp=cmcsc10 scaled\magstep1

\ifamsfonts
   
\else
   
\fi
\newfam\cpfam
\font\fourteenss=cmss12 scaled\magstep1
\font\twelvess=cmss12
\font\tenss=cmss10
\font\niness=cmss9

\font\sevenss=cmss8 at 7pt
\font\sixss=cmss8 at 6pt
\newfam\ssfam
\newfam\msafam \newfam\msbfam \newfam\eufam
\ifamsfonts
 \font\fourteenmsa=msam10 scaled\magstep2
 \font\twelvemsa=msam10 scaled\magstep1
 \font\tenmsa=msam10
 \font\ninemsa=msam9
 \font\sevenmsa=msam7
 \font\sixmsa=msam6
 \font\fourteenmsb=msbm10 scaled\magstep2
 \font\twelvemsb=msbm10 scaled\magstep1
 \font\tenmsb=msbm10
 \font\ninemsb=msbm9
 \font\sevenmsb=msbm7
 \font\sixmsb=msbm6
 \font\fourteeneu=eufm10 scaled\magstep2
 \font\twelveeu=eufm10 scaled\magstep1
 \font\teneu=eufm10
 \font\nineeu=eufm9
 
 \font\seveneu=eufm7
 \font\sixeu=eufm6
 \def\hexnumber@#1{\ifnum#1<10 \number#1\else
  \ifnum#1=10 A\else\ifnum#1=11 B\else\ifnum#1=12 C\else
  \ifnum#1=13 D\else\ifnum#1=14 E\else\ifnum#1=15 F\fi\fi\fi\fi\fi\fi\fi}
 \def\hexmsa{\hexnumber@\msafam}
 \def\hexmsb{\hexnumber@\msbfam} 
\fi
\newdimen\b@gheight             \b@gheight=12pt
\newcount\f@ntkey               \f@ntkey=0
\def\f@m{\afterassignment\samef@nt\f@ntkey=}
\def\samef@nt{\fam=\f@ntkey \the\textfont\f@ntkey\relax}
\def\rm{\f@m0 }
\def\mit{\f@m1 }
\def\cal{\f@m2 }
\def\it{\f@m\itfam}
\def\sl{\f@m\slfam}
\def\bf{\f@m\bffam}
\def\tt{\f@m\ttfam}
\def\caps{\f@m\cpfam}
\def\ssf{\f@m\ssfam}
\ifamsfonts
 \def\msa{\f@m\msafam}
 \def\msb{\f@m\msbfam} \let\bb=\msb
 \def\eu{\f@m\eufam}
\else
 \let \bb=\bf \let\eu=\bf
\fi
\def\fourteenpoint{\relax
    \textfont0=\fourteencp          \scriptfont0=\tenrm
      \scriptscriptfont0=\sevenrm
    \textfont1=\fourteeni           \scriptfont1=\teni
      \scriptscriptfont1=\seveni
    \textfont2=\fourteensy          \scriptfont2=\tensy
      \scriptscriptfont2=\sevensy
    \textfont3=\fourteenex          \scriptfont3=\twelveex
      \scriptscriptfont3=\tenex
    \textfont\itfam=\fourteenit     \scriptfont\itfam=\tenit
    \textfont\slfam=\fourteensl     \scriptfont\slfam=\tensl
      \scriptscriptfont\slfam=\sevensl
    \textfont\bffam=\fourteenbf     \scriptfont\bffam=\tenbf
      \scriptscriptfont\bffam=\sevenbf
    \textfont\ttfam=\fourteentt
    \textfont\cpfam=\fourteencp
    \textfont\ssfam=\fourteenss     \scriptfont\ssfam=\tenss
      \scriptscriptfont\ssfam=\sevenss
    \ifamsfonts
       \textfont\msafam=\fourteenmsa     \scriptfont\msafam=\tenmsa
         \scriptscriptfont\msafam=\sevenmsa
       \textfont\msbfam=\fourteenmsb     \scriptfont\msbfam=\tenmsb
         \scriptscriptfont\msbfam=\sevenmsb
       \textfont\eufam=\fourteeneu     \scriptfont\eufam=\teneu
         \scriptscriptfont\eufam=\seveneu \fi
    \samef@nt
    \b@gheight=14pt
    \setbox\strutbox=\hbox{\vrule height 0.85\b@gheight
                                depth 0.35\b@gheight width\z@ }}
\def\twelvepoint{\relax
    \textfont0=\twelverm          \scriptfont0=\ninerm
      \scriptscriptfont0=\sixrm
    \textfont1=\twelvei           \scriptfont1=\ninei
      \scriptscriptfont1=\sixi
    \textfont2=\twelvesy           \scriptfont2=\ninesy
      \scriptscriptfont2=\sixsy
    \textfont3=\twelveex          \scriptfont3=\ninex
      \scriptscriptfont3=\sixex
    \textfont\itfam=\twelveit    %\scriptfont\itfam=\nineit
    \textfont\slfam=\twelvesl    %\scriptfont\slfam=\ninesl
      \scriptscriptfont\slfam=\sixsl
    \textfont\bffam=\twelvebf     \scriptfont\bffam=\ninebf
      \scriptscriptfont\bffam=\sixbf
    \textfont\ttfam=\twelvett
    \textfont\cpfam=\twelvecp
    \textfont\ssfam=\twelvess     \scriptfont\ssfam=\niness
      \scriptscriptfont\ssfam=\sixss
    \ifamsfonts
       \textfont\msafam=\twelvemsa     \scriptfont\msafam=\ninemsa
         \scriptscriptfont\msafam=\sixmsa
       \textfont\msbfam=\twelvemsb     \scriptfont\msbfam=\ninemsb
         \scriptscriptfont\msbfam=\sixmsb
       \textfont\eufam=\twelveeu     \scriptfont\eufam=\nineeu
         \scriptscriptfont\eufam=\sixeu \fi
    \samef@nt
    \b@gheight=12pt
    \setbox\strutbox=\hbox{\vrule height 0.85\b@gheight
                                depth 0.35\b@gheight width\z@ }}
\twelvepoint
%
%%%%%%%%%%%%%%%%%
%%%Basic skips%%%
%%%%%%%%%%%%%%%%%
%
\baselineskip = 15pt plus 0.2pt minus 0.1pt %was 20pt ...
\lineskip = 1.5pt plus 0.1pt minus 0.1pt
\lineskiplimit = 1.5pt
\parskip = 6pt plus 2pt minus 1pt
\interlinepenalty=50
\interfootnotelinepenalty=5000
\predisplaypenalty=9000
\postdisplaypenalty=500
\hfuzz=1pt
\vfuzz=0.2pt
\dimen\footins=24 truecm % 8 truein in SB
\ifafour
 \hsize=16cm \vsize=22cm
\else
 \hsize=6.5in \vsize=9in
\fi
%
%%%%%%%%%%%%%%%
%%%Footnotes%%%
%%%%%%%%%%%%%%%
%
\skip\footins=\medskipamount
\newcount\fnotenumber
\def\clearfnotenumber{\fnotenumber=0} \clearfnotenumber
\def\fnote{\global\advance\fnotenumber by1 \generatefootsymbol
 \footnote{$^{\footsymbol}$}}
\def\fd@f#1 {\xdef\footsymbol{\mathchar"#1 }}
\def\generatefootsymbol{\iffrontpage\ifcase\fnotenumber
\or \fd@f 279 \or \fd@f 27A \or \fd@f 278 \or \fd@f 27B
\else  \fd@f 13F \fi
\else\xdef\footsymbol{\the\fnotenumber}\fi}
%
%%%%%%%%%%%%%%%%%%%%%%%%%%%%%
%%%Sections and Appendices%%%
%%%%%%%%%%%%%%%%%%%%%%%%%%%%%
%
\newcount\secnumber \newcount\appnumber
\def\clearappnumber{\appnumber=64} \def\clearsecnumber{\secnumber=0}
\clearsecnumber \clearappnumber
\newif\ifs@c % this is true if within a section as opposed to an appendix
\newif\ifs@cd % this is true if the article is being section'd
\s@cdtrue % this is the default
\def\unsectioned{\s@cdfalse\let\section=\subsection}
\newskip\sectionskip         \sectionskip=\medskipamount
\newskip\headskip            \headskip=8pt plus 3pt minus 3pt
\newdimen\sectionminspace    \sectionminspace=10pc
\def\Titlestyle#1{\par\begingroup \interlinepenalty=9999
     \leftskip=0.02\hsize plus 0.23\hsize minus 0.02\hsize
     \rightskip=\leftskip \parfillskip=0pt
     \advance\baselineskip by 0.5\baselineskip%this is a test...
     \hyphenpenalty=9000 \exhyphenpenalty=9000
     \tolerance=9999 \pretolerance=9000
     \spaceskip=0.333em \xspaceskip=0.5em
     \fourteenpoint
  \noindent #1\par\endgroup }
\def\titlestyle#1{\par\begingroup \interlinepenalty=9999
     \leftskip=0.02\hsize plus 0.23\hsize minus 0.02\hsize
     \rightskip=\leftskip \parfillskip=0pt
     \hyphenpenalty=9000 \exhyphenpenalty=9000
     \tolerance=9999 \pretolerance=9000
     \spaceskip=0.333em \xspaceskip=0.5em
     \fourteenpoint
   \noindent #1\par\endgroup }
\def\spacecheck#1{\dimen@=\pagegoal\advance\dimen@ by -\pagetotal
   \ifdim\dimen@<#1 \ifdim\dimen@>0pt \vfil\break \fi\fi}
\def\section#1{\cleareqnumber \s@ctrue \global\advance\secnumber by1
   \par \ifnum\the\lastpenalty=30000\else
   \penalty-200\vskip\sectionskip \spacecheck\sectionminspace\fi
   \noindent {\caps\enspace\S\the\secnumber\quad #1}\par
   \nobreak\vskip\headskip \penalty 30000 }
\def\undertext#1{\vtop{\hbox{#1}\kern 1pt \hrule}}
\def\subsection#1{\par
   \ifnum\the\lastpenalty=30000\else \penalty-100\smallskip
   \spacecheck\sectionminspace\fi
   \noindent\undertext{#1}\enspace \vadjust{\penalty5000}}

\def\appendix#1{\cleareqnumber \s@cfalse \global\advance\appnumber by1
   \par \ifnum\the\lastpenalty=30000\else
   \penalty-200\vskip\sectionskip \spacecheck\sectionminspace\fi
   \noindent {\caps\enspace Appendix \char\the\appnumber\quad #1}\par
   \nobreak\vskip\headskip \penalty 30000 }
\def\ack{\par\penalty-100\medskip \spacecheck\sectionminspace
   \line{\fourteencp\hfil ACKNOWLEDGEMENTS\hfil}%
\nobreak\vskip\headskip }
\def\refs{\begingroup \par\penalty-100\medskip \spacecheck\sectionminspace
   \line{\fourteencp\hfil REFERENCES\hfil}%
\nobreak\vskip\headskip \frenchspacing }
\def\endrefs{\par\endgroup}
%--- Note added
%
%%%%%%%%%%%%%%%%%%%%%%%%%%%%%%%%%
%%%Running heads and footlines%%%
%%%%%%%%%%%%%%%%%%%%%%%%%%%%%%%%%
%
\newif\iffrontpage \frontpagefalse
\headline={\hfil}
\footline={\iffrontpage\hfil\else \hss\twelverm
-- \folio\ --\hss \fi }
%
%%%%%%%%%%%%%%%%
%%%Title page%%%
%%%%%%%%%%%%%%%%
%
\newskip\frontpageskip \frontpageskip=12pt plus .5fil minus 2pt
\def\titlepage{\global\frontpagetrue\hrule height\z@ \relax
               \pubblock\relax }
\def\endtitlepage{\vfil\break\clearfnotenumber\frontpagefalse}
\def\title#1{\vskip\frontpageskip\Titlestyle{\caps #1}\vskip3\headskip}
\def\author#1{\vskip.5\frontpageskip\titlestyle{\caps #1}\nobreak}
\def\and{\par\kern 5pt \centerline{\sl and}}
\def\andauthor{\vskip.5\frontpageskip\centerline{and}\author}

\def\address#1{\par\kern 5pt\titlestyle{\it #1}}
\def\andaddress{\par\kern 5pt \centerline{\sl and} \address}

\def\abstract#1{\par\dimen@=\prevdepth \hrule height\z@ \prevdepth=\dimen@
   \vskip\frontpageskip\spacecheck\sectionminspace
   \centerline{\fourteencp ABSTRACT}\vskip\headskip
   {\noindent #1}}

\def\email#1{\fnote{\tentt e-mail: #1\hfill}}

%
%%%%%%%%%%%%%%%%%%%%
%%%some addresses%%%
%%%%%%%%%%%%%%%%%%%%
%

%

%
\def\QMW{\address{%
   Department of Physics, Queen Mary and Westfield College\break
   Mile End Road, London E1 4NS, UK}}
\def\MONTP{\address{%
    Laboratoire de Physique Math\'ematique\break
    Universit\'e de Montpellier II, Place Eug\`ene Bataillon\break
    34095 Montpellier, CEDEX 5, FRANCE}}
%
%%%%%%%%%%%%%%%%
%%%References%%%
%%%%%%%%%%%%%%%%
%
\newcount\refnumber \def\clearrefnumber{\refnumber=0}  \clearrefnumber
\newwrite\R@fs                              %This opens a file .refs with
\immediate\openout\R@fs=\jobname.refs %the references in order of
                                            %appearance.
\def\closerefs{\immediate\closeout\R@fs} %close file so that TeX can read it
\def\refsout{\closerefs\refs
\unlockat
\input\jobname.refs
\lockat
\endrefs}
\def\refitem#1{\item{{\bf #1}}}%just bolds it so that \bf does not expand
\def\ifundefined#1{\expandafter\ifx\csname#1\endcsname\relax}
\def\[#1]{\ifundefined{#1R@FNO}%
\global\advance\refnumber by1%
\expandafter\xdef\csname#1R@FNO\endcsname{[\the\refnumber]}%
\immediate\write\R@fs{\noexpand\refitem{\csname#1R@FNO\endcsname}%
\noexpand\csname#1R@F\endcsname}\fi{\bf \csname#1R@FNO\endcsname}}
\def\refdef[#1]#2{\expandafter\gdef\csname#1R@F\endcsname{{#2}}}
%
%%%%%%%%%%%%%%%
%%%Equations%%%
%%%%%%%%%%%%%%%
%
\newcount\eqnumber \def\cleareqnumber{\eqnumber=0}
\newif\ifal@gn \al@gnfalse  % this is true if within an \eqalignno
\def\veqnalign#1{\al@gntrue \vbox{\eqalignno{#1}} \al@gnfalse}
\def\eqnalign#1{\al@gntrue \eqalignno{#1} \al@gnfalse}
\def\(#1){\relax%
\ifundefined{#1@Q}
 \global\advance\eqnumber by1
 \ifs@cd
  \ifs@c
   \expandafter\xdef\csname#1@Q\endcsname{{%
\noexpand\rm(\the\secnumber .\the\eqnumber)}}
  \else
   \expandafter\xdef\csname#1@Q\endcsname{{%
\noexpand\rm(\char\the\appnumber .\the\eqnumber)}}
  \fi
 \else
  \expandafter\xdef\csname#1@Q\endcsname{{\noexpand\rm(\the\eqnumber)}}
 \fi
 \ifal@gn
    & \csname#1@Q\endcsname
 \else
    \eqno \csname#1@Q\endcsname
 \fi
\else%
\csname#1@Q\endcsname\fi\global\let\@Q=\relax}
%
%%%%%%%%%%%%%%%%%
%%%Mathematica%%%
%%%%%%%%%%%%%%%%%
%
\newif\ifm@thstyle \m@thstylefalse
\def\mathstyle{\m@thstyletrue}
\def\proclaim#1#2\par{\smallbreak\begingroup%        small --> med???
\advance\baselineskip by -0.25\baselineskip%
\advance\belowdisplayskip by -0.35\belowdisplayskip%
\advance\abovedisplayskip by -0.35\abovedisplayskip%
    \noindent{\caps#1.\enspace}{#2}\par\endgroup%
\smallbreak}%--- defs, thms, ...                     small --> med???
\def\m@kem@th<#1>#2#3{%
\ifm@thstyle \global\advance\eqnumber by1
 \ifs@cd
  \ifs@c
   \expandafter\xdef\csname#1\endcsname{{%
\noexpand #2\ \the\secnumber .\the\eqnumber}}
  \else
   \expandafter\xdef\csname#1\endcsname{{%
\noexpand #2\ \char\the\appnumber .\the\eqnumber}}
  \fi
 \else
  \expandafter\xdef\csname#1\endcsname{{\noexpand #2\ \the\eqnumber}}
 \fi
 \proclaim{\csname#1\endcsname}{#3}
\else
 \proclaim{#2}{#3}
\fi}
\def\Thm<#1>#2{\m@kem@th<#1M@TH>{Theorem}{\sl#2}}%--- Theorem
\def\Prop<#1>#2{\m@kem@th<#1M@TH>{Proposition}{\sl#2}}%--- Proposition
\def\Def<#1>#2{\m@kem@th<#1M@TH>{Definition}{\rm#2}}%--- Definition
\def\Lem<#1>#2{\m@kem@th<#1M@TH>{Lemma}{\sl#2}}%--- Lemma
\def\Cor<#1>#2{\m@kem@th<#1M@TH>{Corollary}{\sl#2}}%--- Corollary
\def\Conj<#1>#2{\m@kem@th<#1M@TH>{Conjecture}{\sl#2}}%--- Conjecture
\def\Rmk<#1>#2{\m@kem@th<#1M@TH>{Remark}{\rm#2}}%--- Remark
\def\Exm<#1>#2{\m@kem@th<#1M@TH>{Example}{\rm#2}}%--- Example
\def\Qry<#1>#2{\m@kem@th<#1M@TH>{Query}{\it#2}}%--- Query
%
%--- Proof
%

%
\def\<#1>{\csname#1M@TH\endcsname}
%
%%%%%%%%%%%%%%%%%%%
%%%Abbreviations%%%
%%%%%%%%%%%%%%%%%%%
%
\def\ref#1{{\bf [#1]}}%--- [ref]
%--- et al.
\def\ie{{\it i.e.\/}}%--- i.e.
%--- e.g.
%--- Cf.
%--- cf.
 %--- double left quote
%--- th as in fifth
\def\nl{\hfil\break}%--- new line
%
%%%%%%%%%%%%%%%%%
%%%Mathematics%%%
%%%%%%%%%%%%%%%%%
%
%--- def over =
%--- Halmos Q.E.D.

%--- implies
%--- is implied by
%--- if and only if
\def\lapprox{\hbox{\lower3pt\hbox{$\buildrel<\over\sim$}}}% approx lt
\def\gapprox{\hbox{\lower3pt\hbox{$\buildrel<\over\sim$}}}% approx gt
\def\quotient#1#2{#1/\lower0pt\hbox{${#2}$}}%--- factor objects
\def\fr#1/#2{\mathord{\hbox{${#1}\over{#2}$}}}
\ifamsfonts
 \mathchardef\empty="0\hexmsb3F %--- better empty set than \emptyset
 \mathchardef\lsemidir="2\hexmsb6E % semidirect |x
 \mathchardef\rsemidir="2\hexmsb6F % semidirect x|
\else
 \let\empty=\emptyset
 \def\lsemidir{\mathbin{\hbox{\hskip2pt\vrule height 5.7pt depth -.3pt
    width .25pt\hskip-2pt$\times$}}}
 \def\rsemidir{\mathbin{\hbox{$\times$\hskip-2pt\vrule height 5.7pt
    depth -.3pt width .25pt\hskip2pt}}}
\fi
%
%--- injective map
%--- surjective map
%--- bijective map
%--- mapping
%--- long mapping
%--- isom over -->
\def\lra{\leftrightarrow}%--- just an abbrev.
%

%
 %--- commutative diagram macro
 %--- map in complex
%
\def\reals{\mathord{\bb R}} %--- reals
 %--- complex nos.
 %--- quaternions
 %--- integers
 %--- rationals
 %--- naturals
 %--- ground field
%
%--- Hom(omorphisms)
%--- tr(ace)
%--- Tr(ace)
%--- End(omorphisms)
%--- Mor(phisms)
%--- Aut(omorphisms)
%--- aut(omorphisms)
%--- supertrace
%--- superdeterminant
%--- kernel
%--- cokernel
%--- image
%
\def\underrightarrow#1{\vtop{\ialign{##\crcr
      $\hfil\displaystyle{#1}\hfil$\crcr
      \noalign{\kern-\p@\nointerlineskip}
      \rightarrowfill\crcr}}} %--- modification of \overrightarrow
\def\underleftarrow#1{\vtop{\ialign{##\crcr
      $\hfil\displaystyle{#1}\hfil$\crcr
      \noalign{\kern-\p@\nointerlineskip}
      \leftarrowfill\crcr}}}  %--- modification of \overleftarrow

\def\comm#1#2{\left[#1\, ,\,#2\right]}%--- [ , ]
%--- { , }
%--- [ , }
%
%--- Lie derivative
%--- vartnl derivative
%--- partial derivative
\def\der#1#2{{{d #1}\over {d #2}}}%--- full derivative
%
%%%%%%%%%%%%%%
%%%Journals%%%
%%%%%%%%%%%%%%
%

\def\NPB#1#2#3{{\sl Nucl. Phys.} {\bf B#1} (#2) #3}

\def\CMP#1#2#3{{\sl Comm. Math. Phys.} {\bf #1} (#2) #3}

\def\PLB#1#2#3{{\sl Phys. Lett.} {\bf #1B} (#2) #3}

\def\FAaIA#1#2#3{{\sl Functional Analysis and Its Application} {\bf #1} (#2)
#3}

\def\TMP#1#2#3{{\sl Theor. Mat. Phys.} {\bf #1} (#2) #3}

\def\JETPL#1#2#3{{\sl  Sov. Phys. JETP Lett.} {\bf #1} (#2) #3}

\lockat
%
%   These are the local macros for the Wtensors paper
%
\def\W{\mathord{\ssf W}}

\def\fr#1/#2{\mathord{\hbox{${#1}\over{#2}$}}}

\def\ket|#1>{\mathord{\vert{#1}\rangle}}

\def\ope#1#2{{{#2}\over{\ifnum#1=1 {z-w} \else {(z-w)^{#1}}\fi}}}

\def\corr<#1>{\mathord{\langle #1 \rangle}}

%
%   these are the references for the paper
%
\refdef[Zam]{
A. B. Zamolodchikov, \TMP{65}{1986}{1205}.}
\refdef[intgeo]{
K. Schoutens, A. Sevrin and P. van Nieuwenhuizen, \NPB{349}{1991}{791},\nl
C.M. Hull, \CMP{156}{1993}{245},\nl
J. De Boer and J. Goeree, \NPB{401}{1993}{369}, ({\tt
hep-th/9206098})\nl
S. Govindarajan and T. Jayaraman, {\sl ``A proposal for the
geometry of $W_n$ gravity''}, ({\tt hep-th/9405146}).}
\refdef[BoWatts]{P. Bowcock and G.M.T. Watts, \NPB{379}{1992}{63}
({\tt hep-th/9111062})).}
\refdef[extgeo]{G. Sotkov and M. Stanishkov, \NPB{356}{1991}{439};
G. Sotkov, M. Stanishkov and C.J. Zhu, \NPB{356}{1991}{245}.\nl
J.L. Gervais and Y. Matsuo, \PLB{274}{1992}{309} ({\tt hep-th/9110028});
\CMP{152}{1993}{317} ({\tt hep-th/9201026}).\nl
J.M. Figueroa-O'Farrill, E. Ramos and S. Stanciu,
\PLB{297}{1992}{289},
({\tt hep-th/9209002}).\nl
W.S. l'Yi and O.A. Soloviev, \NPB{397}{1993}{417}.\nl
J. Gomis, J. Herrero, K. Kamimura and J. Roca, \PLB{339}{1994}{59}
({\tt hep-th/9409024}).}
\refdef[Radul]{A.O. Radul, \JETPL{50}{1989}{371}; \FAaIA{25}{1991}{25}.}
\def\dddot#1{\hbox{$\mathop{#1}\limits^{\ldots}$}}

\def\x{{\bf x}}
\def\vv{{\bf v}}
\def\sss#1{{\scriptscriptstyle{#1}}}
\def\L{{\ssf L}}
\unsectioned
\overfullrule=0pt
\def\pubblock{ \line{\hfil\rm PM/95-33}
               \line{\hfil\rm QMW--PH--95--21}
               \line{\hfil\tt hep-th/9506088}
               \line{\hfil\rm June 1995}}
\titlepage
\title{On $\W_3$-morphisms and the Geometry of Plane Curves}
\author{Eduardo Ramos\email{ramos@lpm.univ-montp2.fr}}
\MONTP
\andauthor{Jaume Roca\email{J.Roca@qmw.ac.uk}}
\QMW
\abstract{We provide a description of $\W_3$ transformations in terms
of deformations of convex curves in two dimensional Euclidean space.
This geometrical interpretation sheds some light on the nature of
finite $\W_3$-morphisms.
We also comment on how this construction can be extended to the
case of $\W_n$ and ``nicely curved'' curves in $\reals^{n-1}$.}

\endtitlepage

$\W$-algebras were first introduced by Zamolodchikov 
in the framework of two dimensional conformal field theory.
It was shown in \[Zam] by 
using the bootstrap method, that the 
extension of the Virasoro algebra
by a single field of spin 3 ($W$) yielded a non-linear associative algebra,
denoted since then by $\W_3$.
The geometrical significance of the Virasoro generator $T$ is well
understood:
$$
Q_{\epsilon}=\oint dz \,\epsilon (z)T(z)
$$
is the generator of conformal transformations (or, equivalently, of
diffeomorphisms of the circle). It is therefore natural
to ask what is the geometrical
significance, if any, of the transformation generated by
$$Q_{\eta}=\oint dz\,\eta(z)W(z)~?
$$

At present, there are several different ways of providing a geometrical
interpretation for these $\W_3$ transformations ($\W_3$-morphisms)
in their classical
limit, {\ie} operator product expansions, or equivalently commutators,
are substituted by Poisson brackets \[BoWatts].
Most of them rely on the extrinsic geometry of curves and
surfaces \[extgeo], although different approaches based on intrinsic
geometry have also proved fruitful \[intgeo].

The purpose of
this note is to give yet another interpretation of $\W_3$-morphisms
in terms of the extrinsic geometry of curves in a particularly simple
and, we believe, enlightening setting.

Our main result can be summarized as follows: {\sl A $\W_3$-morphism
can be defined as a map between strictly convex curves in $\reals^2$.}

The remainder of the paper will be dedicated to justifying the above
statement. Before proceding, however,
it will be necessary to introduce some
well-known results on the geometry of plane curves.

\vskip 0.5truecm

\subsection{The geometry of closed plane curves}
\vskip 0.2truecm
Our basic object of study will be
the geometry of parametrized closed curves
$$\eqalign{\gamma : [a,b]&\rightarrow \reals ^2\cr
t\;\;&\mapsto \x (t),}$$
which are immersions, {\ie} which
satisfy ${\dot \x} (t) = d\x /dt\neq 0$ for all $t$.
In what follows, it will be useful to consider arc-length parametrized
curves. The arc-length function $s: [a,b]\rightarrow \reals$ is defined
as usual by
$$s(t) =\int^t_a |\dot \x (t')| dt'.$$

A tangent vector field to $\gamma$ for all $t$ is given by its velocity
vector.
Its modulus $e$ is the one-dimensional induced {\it einbein}. We can now
introduce the normalized tangent vector $\vv_1$, where
$$
\vv_1 = \der{\x}{s} = {\dot \x \over e}. 
$$
The curvature function can now be defined as the modulus of its
derivative,
$$
\der{\vv_1}{s} = \kappa \vv_2 = {\ddot\x_\perp \over e^2},
$$
where the unit vector $\vv_2$ is by construction orthogonal to $\vv_1$.
We can still go a little a further and define the signed curvature
$\tilde\kappa$ using the natural orientation in $\reals^2$, {\ie}
$\tilde\kappa =\kappa$ if the frame $({\bf v_1},{\bf v_2})$
is positively oriented, and $\tilde\kappa=-\kappa$ otherwise.

%The curvature function can now be defined.
%Since ${\bf v_1}\cdot {\bf v_1} =1$ it follows that
%$$0= {1\over 2}\der{\ }{s} ({\bf v_1}\cdot{\bf v_1})=
%{\bf v_1}\cdot\der{{\bf v_1}}{s},$$
%and therefore
%$$\der{\vv_1}{s} = {1\over e^2}\ddot \x_{\perp}$$
%$is a vector everywhere perpendicular to ${\bf v_1}$. The
%curvature function $\kappa$ is then given by its modulus.
%If we now define $\vv_2$ as the unit vector in the direction of
%$\ddot\x_{\perp}$ we then have the Frenet equation
%$$\der{\vv_1}{s} = \kappa \vv_2.$$

The geometrical relevance of the induced metric and the signed
curvature function is revealed by the fact that any two parametrized
curves in $\reals^2$  with the same values of $e$ and $\tilde\kappa$
should be related by an Euclidean motion,
{\ie} translation and rotation.

A curve is called simple if it is one-to-one. Among the simple closed
curves we may distinguish the convex ones. They are defined by the
property that they are the boundary of a convex set\fnote{ Any
subset of $\reals^2$ is called convex if the line segment joining
any two points in the set belongs to that set.}.
But they are more easily characterized for our purposes by the fact
that they are simple closed curves with $\kappa\geq 0$. We will
work with ``strictly convex'' curves for which the
curvature function is everywhere strictly greater than zero (these
curves are also known as ovals in the mathematical literature).
\vskip0.5 truecm

\subsection{Infinitesimal deformations of strictly convex
curves and the $\W_3$-algebra}
\vskip 0.2truecm
An infinitesimal transformation of a curve $\gamma$ is completely
determined by a vector field with support on
$\gamma$. The importance of working
with strictly convex curves comes from the fact that
$\dot{\bf x}$ and $\ddot{\bf x}_{\perp}$ provide
us with an orthogonal frame for the tangent space
$T_p\reals^2$ for all $p\in\gamma$.
Explicitly,
$$\delta_{\alpha ,\eta}\x = \alpha (t) \dot \x +
\eta (t) \ddot \x_{\perp}.$$

In order to study how the above transformations act on $\gamma$ it is
enough to study their action on $e$ and $\kappa$. They are given by
$$\eqnalign{
\delta_{\alpha ,\eta} e &= \dot\alpha e +\alpha\dot e  -
\eta \kappa^2 e^3 ,\cr
\delta_{\alpha ,\eta}\kappa &= \alpha\dot\kappa
+\ddot\eta\kappa + \dot\eta\left(3{\dot e\kappa\over e} +
2\dot\kappa\right) +\eta\left(\ddot\kappa + 3{\dot e\dot\kappa\over
e} +2{\ddot e\kappa\over e} + \kappa^3 e^2\right),\cr
}$$
where we have used the two-dimensional trivial identity
$$\dddot{\bf x} = {(\dddot{\bf x} \ddot{\bf x}_{\perp})\over
{\ddot{\bf x}_{\perp}^2}} \ddot{\bf x}_{\perp}
+ {(\dddot{\bf x} \dot{\bf x})\over {\dot{\bf x}^2}}\dot{\bf x},
$$
which can also be written in terms of $e$ and $\kappa$ as
$$\dddot\x = \left( {\dot\kappa\over\kappa} + 3{\dot e\over e}\right)
\ddot\x_{\perp} + \left({\ddot e\over e} - \kappa^2 e^2\right)
\dot\x .$$

If we now compute the commutator of two such transformations we find
that, in general, they close among themselves with structure constants
that are
functions of $e$ and $\kappa$. It is possible however to choose a
parametrization such that the algebra has a particularly simple form.
Let us consider
$$\eqnalign{\delta_{\epsilon}\x &=\epsilon\dot\x ,\cr
\delta_{\rho}\x &= \rho \ddot\x_{\perp} -
\left( {1\over 2}\dot\rho +\rho\left( {\dot e\over e} +{2\over 3}
{\dot\kappa\over\kappa}\right)\right)\dot\x  .\cr}$$
A long but straightforward computation now yields
$$\eqalign{
\comm{\delta_{\epsilon_1}}{\delta_{\epsilon_2}} = & \;
\delta_\epsilon,\quad{\rm where}\quad
\epsilon=\epsilon_1\dot\epsilon_2 -\dot\epsilon_1\epsilon_2,
\cr
\comm{\delta_{\epsilon}}{\delta_{\rho}} = & \;
\delta_{\tilde\rho},\quad{\rm where}\quad
\tilde\rho=\epsilon\dot\rho - 2\dot\epsilon\rho,
\cr
\comm{\delta_{\rho_1}}{\delta_{\rho_2}} = & \;
\delta_\epsilon,\quad{\rm where}\quad
\epsilon={{2\over 3}\rho_1\dot\rho_2 T -{1\over 4}\dot\rho_1
\ddot\rho_2 +{1\over 6}\rho_1\dddot\rho_2 - (\rho_1\lra\rho_2)},\cr
}$$
which as we will see is nothing but the algebra of infinitesimal
$\W_3$ transformations
with ``energy-momentum tensor" $T$ given by
$$T= 2{\ddot e\over e}-3{\dot e^2\over e^2}
   -{\dot e\dot\kappa\over e\kappa}
   + {\ddot\kappa\over\kappa} - {4\over 3}{\dot\kappa^2
     \over\kappa^2} +  e^2\kappa^2.$$

In order that the
infinitesimal transformations close when acting on $T$
we have to introduce its ``$\W$-partner"

$$\eqalign{W =& - {1\over 6}{\dddot\kappa\over\kappa}
     + {5\over 6}{\dot\kappa\ddot\kappa\over\kappa^2}
     - {20\over 27}{\dot\kappa^3\over\kappa^3}
     - {2\over 3}\kappa\dot\kappa e^2
\cr
   & -{5\over 6}{\dot\kappa^2\dot e\over\kappa^2 e}
     -{1\over 2}{\dot\kappa\dot e^2\over\kappa e^2}
     +{1\over 2}{\ddot\kappa\dot e\over\kappa e}
     +{1\over 6}{\dot\kappa\ddot e\over\kappa e}.}
$$
Then the infinitesimal transformations on $T$ and $W$ read
$$\eqalign{
\delta_{\epsilon}T = & \; 2 \dddot\epsilon + 2 \dot\epsilon T
+\epsilon\dot T,\cr
\delta_{\epsilon}W = & \; 3\dot\epsilon W +\epsilon\dot W,\cr
\delta_{\rho} T = &\; 3 \dot\rho  W + 2\rho \dot W,\cr
\delta_{\rho} W = & \; -{1\over 6} \rho^{({\rm v})}
- {5\over 6} \dddot\rho
T -{5\over 4}\ddot\rho\dot T - \dot\rho ({3\over 4}\ddot T
+{2\over 3}T^2) -\rho ({1\over 6} \dddot T +{2\over 3} T\dot T)\cr
}$$
which are nothing but the standard classical limit of Zamolodchikov's
$\W_3$-algebra \[BoWatts].

We have just shown that infinitesimal $\W_3$-morphisms can be
understood as infinitesimal deformations of strictly convex curves
in $\reals^2$, therefore it follows that a finite $\W_3$ transformation
can be given a natural geometrical realization as a map between two strictly
convex plane curves.

\vskip 0.5truecm
\subsection{$\W_n$ and nicely curved curves in $\reals^{n-1}$}
\vskip 0.2truecm
It should be clear by now how the preceding techniques may be extended
to the case of $\W_n$. 
Any curve $\gamma$ in $\reals^{n-1}$ satisfies the generalized Frenet
equations
$$\eqalign{
\der{\vv_1} s = & \kappa_1 \vv_2,\cr
\der{\vv_j} s = & \kappa_j \vv_{i+1} - \kappa_{j-1}\vv_{j-1},
\quad\quad 1<j<n-1,\cr
\der{\vv_{n-1}} s = & - \kappa_{n-2} \vv_{n-2}.
}$$
The curvature functions $\kappa_j$  and the orthonormal
vectors $\vv_j$ have the following coordinate expressions
$$
\kappa_j = \sqrt{\left(\x^{\sss{(j+1)}}_\perp\right)^2
\over \dot\x^2\left(\x^{\sss{(j)}}_\perp\right)^2},
\quad\quad 
\vv_j = {\x^{\sss{(j)}}_\perp\over
\sqrt{\left(\x^{\sss{(j)}}_\perp\right)^2}},
$$
and  $\x^{\sss{(j)}}_\perp$ is defined recursively as follows:
$$
\x^{\sss{(j)}}_\perp = \x^{\sss{(j)}}
-\sum_{i=1}^{j-1}{\x^{\sss{(j)}}\cdot\x^{\sss{(i)}}_\perp\over
(\x^{\sss{(i)}}_\perp)^2}\;\x^{\sss{(i)}}_\perp.
$$
The condition that  $\x^{\sss{(j)}}_\perp$ be different from zero for
$j=1,...,n-1$ is tantamount to saying that the curve $\gamma$ is
nicely curved in $\reals^{n-1}$, {\ie}  that all curvatures
$\kappa_1,\ldots,\kappa_{n-1}$ are nowhere vanishing functions and
that the vectors $\vv_1,\ldots,\vv_{n-1}$ form an orthonormal basis.

Accordingly we can describe an arbitrary deformation of the curve 
$\gamma$ as
$$\delta_{\bf\eta}\x=\sum_{j=1}^{n-1} \eta_j\x^{(\sss{j})}_\perp.$$
A detailed proof that there is a particular parametrization of such
transformations
that reproduces the $\W_n$-algebra is beyond the scope of this note,
nevertheless we will sketch here the required steps. The main
observation is that the identity
$$
\x^{\sss{(n)}}=
\sum_{j=1}^{n-1}{\x^{\sss{(n)}}\cdot\x^{\sss{(j)}}_\perp\over
(\x^{\sss{(j)}}_\perp)^2}\;\x^{\sss{(j)}}_\perp
$$
can be rewritten in the form $\L\x=0$, with $\L$ a differential
operator of the form
$$\L ={{\rm d}^n\ \over {\rm dt}^n} - \sum_{j=1}^{n-1} u_j
{{\rm d}^j\ \over {\rm dt}^j},$$
by expressing all (uncontracted) vectors $\x^{\sss{(i)}}_\perp$ in terms of
$\x^{\sss{(k)}}$.

Just as in the two-dimensional case, the einbein $e=\sqrt{\dot\x^2}$
and the curvature functions $\kappa_i$ determine a parametrized curve
up to Euclidean motions.
On the other hand, the coefficients $u_j$ are scalar functions made
out of the derivatives of $\x(t)$, so they are insensitive to
Euclidean motions.
It is clear then that their expression is uniquely determined in terms
of $e$ and $\kappa_i$.
In particular, it can be shown that
$$
u_{n-1}={\x^{\sss{(n)}}\x^{\sss{(n-1)}}_\perp\over
(\x^{\sss{(n-1)}}_\perp)^2}
$$
is a total derivative.
Indeed, from the coordinate expression of Frenet equations,
$$
{d\x^{\sss{(i)}}_{\perp}\over dt} = \x^{\sss{(i+1)}}_{\perp} +
\left({d\ \over dt}
{\rm log}\sqrt{(\x^{\sss{(i)}}_{\perp})^2}\right)\x^{\sss{(i)}}_{\perp}
-{ (\x^{\sss{(i)}}_{\perp})^2\over (\x^{\sss{(i-1)}}_{\perp})^2}\;\;
\x^{\sss{(i-1)}}_{\perp},
$$
we can recursively express all $\x^{\sss{(k)}}$ in terms of
the $\x^{\sss{(j)}}_\perp$ with $1<j\leq k$. A direct computation now
yields
$$
u_{n-1}=
\der{}t\left(\sum^{n-1}_{i=1}\log\sqrt{(\x^{\sss{(i)}}_\perp)^2}\right).
$$
Incidentally, the same procedure can be followed to get the explicit
expressions for all other $u_j$ in terms of $\sqrt{(\x^{\sss{(i)}}_\perp)^2}$
and hence in terms of the einbein and the curvatures.

Now, a rescaling of $\x$
$$
\x\;\longrightarrow\;\left(\prod_{i=1}^{n-1}
(\x^{\sss{(i)}})^2\right)^{1 \over 2n}\x
$$
allows us to eliminate this term and re-express $\L$ in the standard form:
$$\L ={{\rm d}^n\ \over {\rm dt}^n} + \sum_{j=0}^{n-2} W_j
{{\rm d}^j\ \over {\rm dt}^j}.$$
It is now possible to use the results of \[Radul], where
it is shown that the algebra
of deformations of $\L$ preserving its form is nothing but a classical
version of $\W_n$.
\vskip 0.5truecm
\ack
E.Ramos is grateful to the H.M.C. programme of the E.U. for financial support.
J.Roca acknowledges a grant from the Spanish Ministry of
Education and the British Council.
\vskip 0.5truecm

\refsout
\bye